\documentclass[12pt,a4paper]{article}
\usepackage{amsmath,amssymb,mathrsfs,framed,esint,slashed}
\usepackage[colorlinks]{hyperref}
\usepackage{color}
\usepackage[all]{xy}

\usepackage{xcolor}

\newcommand{\addedtext}[1]{{\textcolor{blue}{#1}}} 
\newcommand{\changedtext}[1]{{\textcolor{purple}{#1}}}

\newtheorem{theorem}{Theorem}[section]

\newtheorem{remark}[theorem]{Remark}

\newcommand{\rem}[1]{}

\newcommand{\de}{{\rm d}}

\newcommand{\bv}{{\mathbf{v}}}
\newcommand{\bp}{{\mathbf{p}}}

\newcommand{\bX}{{\mathbf{X}}}

\newcommand{\bx}{{\mathbf{x}}}

\newcommand{\bz}{{\mathbf{z}}}

\newcommand{\bA}{{\mathbf{A}}}

\newcommand{\bE}{{\mathbf{E}}}
\newcommand{\bB}{{\mathbf{B}}}
\newcommand{\bJ}{{\mathbf{J}}}

\newcommand{\bu}{{\boldsymbol{u}}}

\newcommand{\beq}{\begin{equation}}
\newcommand{\eeq}{\end{equation}}
\newcommand{\ben}{\begin{eqnarray}}
\newcommand{\een}{\end{eqnarray}}

\def\pdf{f}
\def\WaveFunc{\Psi}

\newcommand\norm[1]{\left|#1\right|}

\def\Lag{\mathscr{L}}
\def\Ham{H}
\def\KvNOp{\widehat{L}_\Ham}

\def\PoissonTensor{\mathbb{J}}

\begin{document}

\title{Koopman wavefunctions and Clebsch variables
\\in Vlasov-Maxwell kinetic  theory\footnote{Contribution to the collection ``Hamiltonian Methods in Plasma Physics'' in the Journal of Plasma Physics.}}
\author{
Cesare Tronci$^{1,2}$ and Ilon Joseph$^{3}$
\\
\footnotesize\it 
$^{1\!}$Department of Mathematics, University of Surrey, Guildford, UK
\\
\footnotesize
\it 
$^{2\!}$Department of Physics and Engineering Physics, Tulane University, New Orleans LA, USA
\\
\footnotesize
\it 
$^{3\!}$Physics Division, Lawrence Livermore National Laboratory, Livermore CA, USA\vspace{-.15cm}
}
\date{}
\maketitle

\begin{abstract} \color{black}Motivated by recent discussions on the possible role of quantum computation in plasma simulations, here we present different approaches to Koopman's Hilbert-space formulation of classical mechanics in the context of Vlasov-Maxwell kinetic theory. The celebrated Koopman-von Neumann construction is provided with two different Hamiltonian structures: one is canonical and recovers the usual Clebsch representation of the Vlasov density, the other  is noncanonical and appears to overcome certain issues emerging in the canonical formalism.
Furthermore, the canonical structure is restored for a variant of the Koopman-von Neumann construction that carries a different phase dynamics. Going back to van Hove's prequantum theory, the corresponding Koopman-van Hove equation provides an alternative Clebsch representation which is then coupled to the electromagnetic fields. Finally, the role of  gauge transformations in the new context is discussed in detail.

\end{abstract}

{
\tableofcontents
}

\section{Introduction}

Ongoing discussions \cite{DOE} on the potential of quantum information and computation in plasma physics have recently led to exploiting Hilbert-space approaches in the numerical simulation of magnetized plasmas \cite{Dodin20,Engel19,Engel20,Joseph20}, of the Navier-Stokes equations \cite{Gaitan20}, and of arbitrary non-Hamiltonian systems of equations \cite{ Joseph20,Liu20}. In particular, recent work \cite{Joseph20} has emphasized the role of Koopman wavefunctions in classical dynamics while their usage in describing hybrid quantum-classical system was presented in \cite{BoGBTr19,boucher,GBTr20,GBTr21b,Sudarshan,Sugny}. In this paper we want to show how these enter in the variational and Hamiltonian formulation of Vlasov kinetic theory. 

In the context of the classical Liouville equation $\partial_t f=\{H,f\}$, the idea of a Koopman wavefunction emerges naturally from the fact that the phase-space density is positive-definite, so that the relation $f(x,p)=|\Psi(x,p)|^2$ induces a Hilbert-space description of classical mechanics hinging on the Koopman-von Neumann (KvN) equation \cite{Koopman,Mauro,VonNeumann2}
\beq
i\hbar\partial_t\Psi=\widehat{L}_H\Psi
\,,\qquad\text{ where }\qquad
\widehat{L}_H:=\{i\hbar H,\cdot\,\}
,
\label{KvN-eq}
\eeq
and $\{\cdot,\cdot\}$ denotes the canonical Poisson bracket.
Here, the Liouvillian operator $\widehat{L}_H$ is self-adjoint so that the Koopman wavefunction undergoes unitary dynamics, thereby leading to a quantum analogy. While the Koopman propagator $\exp(-i\hbar^{-1}\widehat{L}_H)$ (Koopman operator) often appears in the dynamical systems literature  \cite{Mezic,Giannakis}, the role of Koopman wavefunctions is being recognized in physics only recently. Over the decades,  prominent authors \cite{Berry,Chirikov88,Wiener,tHooft} discussed Koopman wavefunctions without referring to Koopman's original work. However, the interest in Koopman wavefunctions has been recently revived by their role in the quantum-classical divide \cite{Bondar}.

One of the important motivations for the introduction of the Koopman formalism is to develop a rigorous Hilbert space theory for the approximation of the particle distribution function (PDF). 
For example, it is well known that the linearized Vlasov-Maxwell equation possesses Case-van Kampen eigenmodes and one might wish to understand how an expansion in eigenmodes, or some other complete set of eigenfunctions, converges to a solution of the nonlinear Vlasov-Maxwell equations. 
{This point of view was taken by Koopman and von Neumann in their development of ergodic theory.}

Yet another important application of the Koopman approach  is the development of quantum algorithms for simulating nonlinear classical dynamical systems  \cite{Joseph20}.
In principle, quantum computers can store and process an exponentially large amount of information efficiently and can perform certain calculations, such as the ``quantum'' Fourier transform, very efficiently \cite{NielsenChuangBook}.
While quantum computers hold great promise, they can only perform linear unitary operations on normalized wavefunctions.
A number of authors \cite{Alanson92,Chirikov88, Joseph20} have reformulated nonlinear non-Hamiltonian dynamics as a linear unitary evolution of functions on phase space.
Thus, as shown in Ref. \cite{Joseph20}, quantum Hamiltonian simulation algorithms for the Koopman evolution operator can achieve up to an exponential speedup over an Eulerian discretization of the PDF and using amplitude estimation of physical observables leads to up to a quadratic speedup over Lagrangian discretizations based on Monte Carlo and particle-in-cell (PIC).

Depending on the  choice of phase, Koopman wavefunctions are related to the Clebsch representation of the Vlasov distribution \cite{Morrison1}, which may be of more common knowledge in the plasma physics community. 
{\color{black}However, as discussed in Section \ref{sec:KvN-canonical}, t}he complete identification between Koopman wavefunctions and Clebsch variables may require the Koopman phase to be singular at some points or regions {\color{black}in} phase space \cite{Joseph20}.
 In turn, the phase is irrelevant in the Koopman-von Neumann theory and one should be able to set it to zero without affecting the general formulation of the theory. {\color{black}In order to overcome this possible issue, we develop an alternative noncanonical formulation that indeed allows for a zero phase. As shown in Section \ref{sec:KvN-Euler-Poincare}, this formulation follows from a suitable adaptation of the Euler-Poincar\'e variational principle for standard Vlasov dynamics \cite{Cendra,Squire,TronciRec}. In this case, the relation to standard Clebsch variables is replaced a momentum map structure that is discussed in Section \ref{sec:KvN-noncanonical}. Then, Section \ref{sec:KvN-Maxwell} applies this construction to the Vlasov-Maxwell system.}

In addition, a variant of the Koopman-von Neumann construction called the Koopman-van Hove equation (KvH) will also be illustrated {\color{black}in Section \ref{sec:KvH_theory}. As shown in Section \ref{sec:KvH_equation}, in this variant the Koopman-von Neumann equation emerges as the amplitude equation, while the evolution of the phase is given by the classical action, i.e. the integral of the Lagrangian function, following Feynman's general prescription. The KvH formulation is based on a slight extension of standard canonical transformations first appeared in van Hove's thesis \cite{VanHove} and briefly discussed in Section \ref{sec:vanHove_transformations}. As the KvH construction is based on a canonical Hamiltonian structure, its relation to the Vlasov density comprises an extension of the standard Clebsch representation, as presented in Section \ref{sec:KvHdensity}. Then, Section \ref{sec:KvH-Maxwell} couples the KvH equation to the evolution of the electromagnetic fields and presents the resulting Hamiltonian structure. Finally, Section \ref{sec:gaugetransf} discusses the role of gauge transformations in the  KvH context.}

\section{
Koopman-von Neumann theory \label{sec:KvN_theory}}

In this section, we  present two alternative formulations of the Koopman-von Neumann (KvN) theory. 
{\color{black} The first formulation is canonical. In this case, the Hamiltonian functional depends on the phase and vanishes if the latter is initially set to zero. Indeed, this Hamiltonian generally differs from the expression of the total energy obtained by Koopman's prescription  $f=\norm{\Psi}^2$. This ambiguity may be overcome by enforcing a specific constraint on the phase \cite{Joseph20} thereby leading to branch-cut singularities. We show that these intricacies can be removed by resorting to an alternative noncanonical formulation that is entirely built upon the prescription  $f=\norm{\Psi}^2$, so that the Hamiltonian functional is independent of the phase which can then be set to zero.}
In order to simplify the notation, most of our discussion will treat the case of a single degree of freedom, so that the phase space is two-dimensional.
The generalization to multiple degrees of freedom is straightforward, and will be used when discussing the Vlasov-Maxwell system.

\subsection{Clebsch variables and canonical structure
\label{sec:KvN-canonical}}
As first discussed in \cite{Dodin1}, the Koopman-von Neumann (KvN) 
equation is Hamiltonian with the canonical Hamiltonian structure arising from the following Dirac-Frenkel variational principle \cite{Frenkel}:
\beq
\delta\int_{t_1}^{t_2}\!\big\langle \Psi,i\hbar\partial_t\Psi-\widehat{L}_H\Psi\big\rangle\,\de t=0
\eeq
Here, $\Psi$ is a square-integrable complex function, $\langle A,B\rangle=\operatorname{Re}\langle A|B\rangle$ is the real-valued pairing associated to the standard $L^2$ inner product $\langle A|B\rangle=\int\! A(x,p)^* B(x,p)\,\de x\de p$,  and we have $\operatorname{Im}\langle A|B\rangle=\langle iA,B\rangle$. As discussed in \cite{BoGBTr19,Joseph20}, this Hamiltonian structure readily identifies a Hamiltonian functional given  by
\[
h_{\text{\tiny KvN}}(\Psi)=\langle\Psi|\widehat{L}_H\Psi\rangle= \hbar\int\! H  \operatorname{Im}\{\Psi^*,\Psi\} \,\de x\de p
\,.
\]
This is accompanied by the usual Poisson bracket from standard quantum mechanics \cite{ChMa,FoHoTr19}:
\beq
\{\!\{h,k\}\!\}(\Psi)=\frac1{2\hbar}\operatorname{Im}\left\langle\frac{\delta h}{\delta \Psi}\bigg|\frac{\delta k}{\delta \Psi}\right\rangle,
\label{CanPB}
\eeq
where we have introduced the double-bracket notation $\{\!\{\cdot,\cdot\}\!\}$ to distinguish from the canonical Poisson bracket.
Then, the Madelung transform $\Psi=\sqrt{D}e^{iS/\hbar}$ expresses the Koopman wavefunction in terms of the density $D$ and the phase $S$, thereby leading to the celebrated Clebsch representation of the real-valued Vlasov density \cite{HoKu,MaWe2,Morrison1}, that is
\beq\label{Clebsch}
f=\hbar\operatorname{Im}\{\Psi^*,\Psi\}=\{D,S\}
. 
\eeq
 Here, the classical Liouville equation $\partial_t f=\{H,f\}$ follows by combining the Jacobi identity with the relations
\beq\label{DSeqs}
\partial_t D=\{H,D\}
\,,\qquad\quad 
\partial_t S=\{H,S\}
\,,
\eeq
which arise from Eq. \eqref{KvN-eq}.
For example, this formulation appeared in \cite{Neiss}, where the Clebsch representation was used together with Koopman's original prescription $f=|\Psi|^2$ in the context of the Vlasov-Poisson system for electrostatic plasmas. 
 In symplectic geometry,  the map $\Psi\mapsto \hbar\operatorname{Im}\{\Psi^*,\Psi\}$ identifies  a momentum map  for the action of canonical transformations on the Hilbert space $\cal H$ of Koopman wavefunctions, which is endowed with the canonical Poisson structure  in  Eq. \eqref{CanPB}. See Section \ref{sec:KvN-noncanonical} later on for  the definition of momentum maps.

This picture leads us to two following important observations. First, the  relation \eqref{Clebsch} differs from  Koopman's prescription $f=|\Psi|^2$ for the phase-space density. Second, the Clebsch representation \eqref{Clebsch} does not generally identify a probability density, since $\int\{D,S\}\,\de x\de p=0$ whenever $D$ and $S$ are differentiable continuous functions. In \cite{Joseph20}, a solution to this apparent issue was provided by selecting a singular  phase ensuring   the following consistency condition:
\beq\label{KvNconsistency} 
|\Psi|^2=\hbar\operatorname{Im}\{\Psi^*,\Psi\}
,\qquad\, \text{ or, equivalently, }\qquad\ 
D=\{D,S\}
\,.
\eeq 
Due to the equations of motion \eqref{DSeqs}, if this relation holds true as an initial condition, then it will hold for all time. In this case, the expression of the phase is necessarily singular and this is strictly necessary to produce the boundary terms  generally required for the Clebsch representation  \eqref{Clebsch} to hold true. 
 Then, in this context, we realize that  the KvN construction may be envisioned as a special case of Clebsch representation of the Vlasov density \cite{Joseph20}.

Notice that this representation requires a non-vanishing phase, which again contrasts with the phase-invariance of the original prescription $f=|\Psi|^2$. 
Indeed, while one might wish 
to set the phase to zero {\color{black} and identify} the Koopman wavefunction with a real-valued amplitude, this possibility is excluded by the Clebsch representation, which requires a nonzero phase by construction.

Whether or not the requirement of a nonzero, possibly singular  phase is seen as a possible issue, it may still be desirable to develop a Hamiltonian structure of KvN theory that does not necessitate the presence of a nontrivial phase.
In view of this,  we will proceed by considering an alternative noncanonical Hamiltonian structure of Koopman-von Neumann theory, which appears to overcome this potential ambiguity.
 
{\color{black}
\begin{remark}[Guillemin-Sternberg collectivization]\label{Collectivization} 
We remark that, in both the canonical and the noncanonical case, Koopman wavefunctions provide a representation of the Liouville (Vlasov) density whose dynamics is then governed by the usual Lie-Poisson bracket on the Poisson algebra of phase-space functions \cite{MaWe1,MaWeRaScSp, Morrison2bis}. The procedure taking the Koopman Hamiltonian structure into the Vlasov Lie-Poisson structure is an example of collectivization by a momentum map \cite{GuSt80}, which serves as the unifying geometric framework in this work; see Section \ref{sec:KvN-noncanonical}. Specifically,  the canonical action of a Lie group on a Poisson manifold induces a momentum map generalizing Noether’s conserved quantity occurring in the particular case of a symmetry group. Then, when a Hamiltonian function(al) can be entirely written in terms of this momentum map, the Hamiltonian is called ‘collective’, borrowing the terminology from nuclear physics \cite{GuSt80}. Here, we derive collective Hamiltonians for a series of models in Vlasov dynamics for which the Lie group is given by canonical transformations (or variants thereof, as in the case of Section \ref{sec:vanHove_transformations}). In this process, different Clebsch-type representations emerge from different actions on the Hilbert space of Koopman wavefunctions. Alternatively, depending on convenience, one may prefer to keep working with the Vlasov density itself.
\end{remark}
}

\subsection{Euler-Poincar\'e 
variational formulation \label{sec:KvN-Euler-Poincare}
}

In this section, we shall start by presenting the Euler-Poincar\'e formulation  of Koopman-von Neumann theory.  
We start from the Euler-Poincar\'e variational principle for the classical Liouville equation:
\beq\label{EPVP1}
\delta\int_{t_1}^{t_2}\int\! f(x,p,t)\big(pu(x,p,t)-H(x,p)\big)\,\de x\de p\,\de t=0
\,.
\eeq
As discussed extensively in \cite{Cendra,Squire,TronciRec}, this variational principle arises from a symmetry reduction from the Lagrangian-path formulation in phase-space to the corresponding Eulerian-variable description. Indeed, if we introduce the notation $\Bbb{J}^{j\ell}=\{z^j,z^{\ell}\}$ and $\bz_0=(x_0,p_0)$ is the phase-space label coordinate, then the equation of motion $\dot{\boldsymbol\eta}=\Bbb{J}\nabla H(\boldsymbol\eta)$ of the Lagrangian trajectory $\boldsymbol\eta(\bz_0,t)=(\eta_q(\bz_0,t),\eta_p(\bz_0,t))$  follows from the variational principle $\delta\int_{t_1}^{t_2}\!\int\! f_0(\eta_p\dot{\eta}_q-H(\eta_q,\eta_p))\,\de x_0\de p_0\,\de t=0$, where $f_0(\bz_0)$ is the reference Liouville density. In turn, the latter variational principle leads to the Eulerian counterpart \eqref{EPVP1} by a mere change of variables.
Upon denoting by $\bz=(x,p)$  the Eulerian phase-space   coordinate, the Eulerian phase-space density evolves according to the Lagrange-to-Euler map:
\beq\label{LtEf}
f(\bz,t)=\int\!f_0(\bz_0)\,\delta(\bz-\boldsymbol\eta(\bz_0,t))\,\de^2 z_0=\frac{f_0(\bz_0)}{\det \nabla\boldsymbol\eta(\bz_0)}\bigg|_{\bz_0=\boldsymbol\eta^{-1}(\bz,t)}
.
\eeq
Here, $\det \nabla\boldsymbol\eta$ denotes the Jacobian determinant. In addition, we define the Eulerian phase-space vector field $\bX=(u,\sigma)$ such that
\beq\label{vectField}
\bX(\bz,t)=\big(u(\bz,t),\sigma(\bz,t))=\dot{\boldsymbol\eta}(\bz_0,t)\big|_{\bz_0=\boldsymbol\eta^{-1}(\bz,t)}
\,.
\eeq
In this setting, we also define the infinitesimal displacement ${\boldsymbol\Xi}(\bz,t)=\delta{\boldsymbol\eta}(\bz_0,t)\big|_{\bz_0=\boldsymbol\eta^{-1}(\bz,t)}$ so that the Euler-Poincar\'e variations read
\beq
\delta\bX=\partial_t\boldsymbol\Xi+\bX\cdot\nabla\boldsymbol\Xi-\boldsymbol\Xi\cdot\nabla\bX
\,,\qquad\qquad
\delta f=-\operatorname{div}(f\boldsymbol\Xi)
\,,
\label{vars1}
\eeq
which are obtained from the defining relations \eqref{LtEf} and \eqref{vectField}. 
Here,  $\boldsymbol\Xi$  is an arbitrary displacement vanishing at the endpoints, i.e. $\boldsymbol\Xi(t_1)=\boldsymbol\Xi(t_2)=0$. 
Then, the Euler-Poincar\'e variational principle \eqref{EPVP1} yields $\bX=\Bbb{J}\nabla H$, which is accompanied by the continuity equation $\partial_t f+{\operatorname{div}(f\bX)=0}$ following from  \eqref{LtEf}, thereby returning $\partial_t f+\{f,H\}=0$.

The Euler-Poincar\'e approach for the Koopman-von Neumann theory arises immediately from the construction above. Indeed, the Koopman description $f=|\Psi|^2$ immediately leads to
\beq\label{HDevol}
\Psi(\bz,t)=\frac{\Psi_0(\bz_0)}{\sqrt{\det \nabla\boldsymbol\eta(\bz_0)}}\bigg|_{\bz_0=\boldsymbol\eta^{-1}(\bz,t)}
,
\eeq
which is the standard Lagrange-to-Euler map for half-densities \cite{BatesWeinstein}. 
This evolution law leads to the relations
\beq\label{psivar}
\partial_t\Psi=-\bX\cdot\nabla\Psi-\frac12\operatorname{div}(\bX)\Psi
\,,\qquad\qquad
\delta\Psi=-\boldsymbol\Xi\cdot\nabla\Psi-\frac12\operatorname{div}(\boldsymbol\Xi)\Psi.
\eeq
One can introduce the operators $\widehat{\Lambda}_\ell=-i\hbar\partial_\ell$ to write the above as Schr\"odinger-like equations:
\[
i\hbar\partial_t\Psi=\frac12\big[\widehat{X}^\ell,\widehat{\Lambda}_\ell\big]_+\Psi
\,,\qquad\qquad
i\hbar\delta\Psi=\frac12\big[\widehat{\Xi}^\ell,\widehat{\Lambda}_\ell\big]_+\Psi
\,,
\]
where $[\cdot,\cdot]_+$ denotes the anticommutator, $\widehat{X}^\ell$ are simply the multiplicative operators identified by the components of $\bX$, and likewise for $\widehat{\Xi}^\ell$. 
Notice that here we are not asking for the vector field $\bX$ to be Hamiltonian: as we shall see, this property will result as a consequence of the variational principle under consideration. In more generality, the  evolution law \eqref{HDevol} and its equation of motion in \eqref{psivar} also appear in other contexts involving non-Hamiltonian dynamical systems  \cite{Alanson92,Joseph20}.

Since we are interested in the Hamiltonian structure, we shall write the Euler-Poincar\'e variational principle in terms of an arbitrary Hamiltonian functional $h(\Psi)$ as follows:
\beq\label{EPKvNVP}
\delta\int_{t_1}^{t_2}\!\left(\int\! |\Psi(x,p,t)|^2\,pu(x,p,t)\,\de x\de p-h(\Psi)\right)\de t=0
\,,
\eeq
\noindent
where we recall \eqref{vectField}. 
Here,  the arbitrary functional $h(\Psi)$ generally depends on $\Psi$ and its conjugate $\Psi^*$.
Upon combining \eqref{psivar} with the first equation in \eqref{vars1}, the Euler-Poincar\'e variational principle yields
\beq\label{Xexp}
\bX=-\frac1{|\Psi|^2}\,\Bbb{J}\!\left(\Psi\diamond\frac{\delta h}{\delta \Psi}\right)
\,,\qquad\qquad
\Psi\diamond\frac{\delta h}{\delta \Psi}:=\frac12
\operatorname{Re}\!\left(\frac{\delta h}{\delta \Psi}^{\!*}\nabla\Psi-\Psi^*\nabla\frac{\delta h}{\delta \Psi}\right)
.
\eeq
{\color{red}Here,} we have used the functional derivative notation so that $
\delta h:=\left\langle{\delta h}/{\delta \Psi},\delta\Psi\right\rangle$
and,  following Ref. \cite{HoMaRa1998}, we have defined the diamond operator $\diamond$ for compactness of notation.

Then, the wavefunction equation for an arbitrary Hamiltonian is obtained by substituting \eqref{Xexp} in the first equation of \eqref{psivar}, which leads to quite a cumbersome explicit form. 
However, in  the case of Koopman-von Neumann theory, this simplifies substantially. 
Indeed, if the Hamiltonian functional depends only on $f=|\Psi|^2$, then the chain rule relation ${\delta h}/{\delta \Psi}={\color{red}2}({\delta h}/{\delta f})\Psi$ transforms  \eqref{Xexp} to the Hamiltonian  vector field
\[
\bX=\Bbb{J}\nabla\frac{\delta h}{\delta f}=:\bX_{{\delta h}/{\delta f}}
\,,
\]
thereby recovering the Koopman-von Neumann equation \eqref{KvN-eq} in the general form
\beq
i\hbar\partial_t\Psi=\widehat{L}_{{\delta h}/{\delta f}}\Psi
\,.
\label{KvNHam}
\eeq
In this case, the Lagrangian trajectory is a canonical transformation so that  the propagator \eqref{HDevol} reduces to $\Psi(\bz,t)=\Psi_0(\bz_0)|_{\bz_0={\boldsymbol\eta}^{-1}(\bz,t)}$.

Notice that, while here we have used canonical coordinates associated to the canonical symplectic form $\de x\wedge \de p$, the possibility of a noncanonical structure does not pose any difficulty. Indeed, one would simply redefine canonical transformations as \emph{symplectomorphisms}, that is smooth invertible transformations preserving the noncanonical symplectic structure. Then, the Liouvillian operator $\widehat{L}_H=\{i\hbar H,\,  \}$ would naturally involve the corresponding noncanonical Poisson bracket. 

\subsection{Non-canonical
Poisson bracket and momentum map structure \label{sec:KvN-noncanonical}}
We realize that equation \eqref{KvNHam} arises naturally from the Euler-Poincar\'e formulation of the Vlasov equation, while its underlying structure is quite involved and this reflects in an intricate noncanonical Hamiltonian structure. Given an arbitrary functional $k(\Psi)$, the usual relation $\dot{k}=\{\!\{k,h\}\!\}$ leads in this case to
\begin{equation}\label{KvNPBNC}
\{\!\{k,h\}\!\}(\Psi)=\int\!\frac1{|\Psi|^2}\left(\Psi\diamond\frac{\delta k}{\delta \Psi}\right)\cdot\Bbb{J}\left(\Psi\diamond\frac{\delta h}{\delta \Psi}\right)\de x\de p
\,,
\end{equation}
which  reduces to the Lie-Poisson bracket 
\begin{equation}\label{VlasovLPB}
\{\!\{k,h\}\!\}(f)=\int \!f\left\{\frac{\delta k}{\delta f},\frac{\delta h}{\delta f}\right\}\de x\,\de p
\end{equation}
in the case of functionals depending only on $f=|\Psi|^2$. 
To obtain \eqref{KvNPBNC}, we have used the relation $\langle{\Psi\diamond\Upsilon},{\bf W}\rangle=\langle\Upsilon,{\bf W}\cdot\nabla\Psi+\operatorname{div}({\bf W})\Psi/2\rangle$ for any Koopman wavefunctions $\Psi$ and $\Upsilon$, and any vector field ${\bf W}$. Indeed, this relation leads to
\[
\dot{k}=\left\langle\frac{\delta k}{\delta \Psi},\partial_t\Psi\right\rangle=
-\left\langle\frac{\delta k}{\delta \Psi},\bX\cdot\nabla\Psi+\frac12\operatorname{div}(\bX)\Psi\right\rangle=
-\left\langle\Psi\diamond\frac{\delta k}{\delta \Psi},\bX\right\rangle
=\{\!\{k,h\}\!\}(\Psi)
\]
where the last equality follows from \eqref{Xexp}.
While expanding the terms in the noncanonical structure \eqref{KvNPBNC} does not lead to much insight, the present formulation has the advantage of {\color{black}restoring}
the use of the relation $f=|\Psi|^2$ {\color{black}everywhere in the problem. For example, the standard expression of the total energy now coincides unambiguously with the Hamiltonian functional, which no longer vanishes upon setting the phase to zero.}

We conclude this section by showing that the map $\Psi\mapsto|\Psi|^2$ is a momentum map \cite{MaRa} associated to the following (left) group action of canonical transformations: $\Psi\mapsto\Psi\circ{\boldsymbol\eta}^{-1}$, where $\circ$ denotes composition of functions. {\color{black}By the arguments in Remark \ref{Collectivization}, this ensures that the Poisson bracket \eqref{KvNPBNC} consistently reduces to \eqref{VlasovLPB} for any two functionals $k$ and $h$ of the type $k(\Psi)=\tilde{k}(|\Psi|^2)$.} Given the action of a Lie group $G$  on a manifold $M$ with  Poisson structure $\{\!\{\cdot,\cdot\}\!\}$, a momentum map ${\cal J}:M\to\mathfrak{g}^*$ is defined as
\beq\label{Momapdef}
\{\!\{k(x),\langle {\cal J}(x),\xi\rangle\}\!\}=\langle \de k(x),\xi_M(x)\rangle
\eeq
for any function $k$ on $M$ and any element $\xi$ of the Lie algebra $\mathfrak{g}$ of $G$. Here, $\mathfrak{g}^*$ denotes the dual space of $\mathfrak{g}$, $\langle\cdot,\cdot\rangle$ is the duality pairing,
 $\de$ is the differential on $M$, and $\xi_M$ denotes the infinitesimal generator of the  $G-$action on $M$. If $G$ is a symmetry of the Hamiltonian, then ${\cal J}(x)$ is conserved in time by Noether's theorem.  
The Lie algebra of the group of canonical transformations is the space of Hamiltonian vector fields $\bX=\bX_\xi$ with $\xi=\xi(x,p)$. Up  to the addition of irrelevant constants, this space can be identified with the space of phase-space functions endowed with the canonical Poisson bracket. Then, the corresponding infinitesimal action on the Koopman Hilbert space $\mathcal{H}=L^2(\Bbb{R}^2)$ is given by $\xi_{\mathcal{H}}(\Psi)=-\bX_\xi\cdot\nabla\Psi=\{\xi,\Psi\}$. 
Then, upon using the $L^2-$pairing, we compute
\[
\Psi\diamond\frac{\delta}{\delta \Psi}\langle |\Psi|^2,\xi\rangle=2\Psi\diamond(\xi\Psi)
=-|\Psi|^2\nabla\xi
\]
so that
\[
\{\!\{k,\langle |\Psi|^2,\xi\rangle\}\!\}
=-
\int\!\frac{\delta k}{\delta \Psi}\{\Psi,\xi\}\de x\de p
\]
thereby proving \eqref{Momapdef}. This picture extends the usual Clebsch representation treated in Section \ref{sec:KvN-canonical} to the noncanonical case. 
By proceeding analogously, one proves that the quantities
\[
\int|\Psi|^2\,\de^3 p
\,,\qquad\qquad
\int p\,|\Psi|^2\,\de^3 p
\]
comprise the \emph{plasma-to-fluid} momentum map \cite{MaWeRaScSp}. Specifically, these are momentum maps for the action of momentum translations $(x,p)\mapsto (x,p-\de \varphi(x))$ and configuration-space diffeomorphisms $(x,p)\mapsto(\eta(x),-p\de\eta(x))$, respectively.

\subsection{The KvN-Maxwell system \label{sec:KvN-Maxwell}}

In this section, we want to apply the previous noncanonical formulation  to the Maxwell-Vlasov system of magnetized plasmas. The canonical treatment is found in \cite{Neiss} for the case of the electrostatic limit. In the full electromagnetic case,  gauge-invariance naturally acquires a prominent role. In particular, 
gauge freedom manifests itself in the Lagrangian formulation of Maxwell's equations through the fact that the Lagrangian is independent of $\partial_t \Phi$, and, hence, variations with respect to $\Phi$ serve  to enforce Gauss' law as a constraint. 
 In the absence of sources, the Maxwell Lagrangian is written as
\[
L_\text{\tiny Max}=-\epsilon_0\int\bE\cdot (\partial_t\bA +\nabla\Phi) \,\de^3 x-h_\text{\tiny Max}(\bA,\bE)
\,,
\]
where 
\beq
h_\text{\tiny Max}(\bA,\bE)=\frac{\epsilon_0}2\|\bE\|^2+ \frac1{2\mu_0}\|\nabla\times\bA\|^2
\label{MaxHam}
\eeq
 is the Maxwell Hamiltonian expressed in terms of the standard $L^2-$norm. If we express the Vlasov density in noncanonical coordinates, we can construct the variational principle by an immediate extension of Eq. \eqref{EPKvNVP}. We write
\begin{multline}\label{KMVP}
\delta\int_{t_1}^{t_2}\!\bigg[\int\!|\Psi|^2\Big((m\bv+q\bA)\cdot\bu-q\Phi\Big)\,\de^3x\,\de^3v-\epsilon_0\int\bE\cdot (\partial_t\bA+\nabla\Phi)  \,\de^3x-h(\Psi,\bA,\bE)
\bigg]\de t=0
\,,
\end{multline}
where 
\begin{align*}
h(\Psi,\bA,\bE)=&\,h_\text{\tiny Max}(\bA,\bE)+\frac{m}2\int\!|\Psi|^2v^2 \,\de^3x\,\de^3v
\\
=&\,
\frac{m}2\int\!|\Psi|^2v^2 \,\de^3x\,\de^3v+ \frac{\epsilon_0}2  \int\! |\bE|^2  \,\de^3x+\frac1{2\mu_0}\int|\nabla\times\bA|^2\,\de^3x
\end{align*} 
is the Hamiltonian, 
 while Gauss' law is enforced in \eqref{KMVP}   by the Lagrange multiplier $\Phi$.
Using \eqref{psivar} and the first equation in \eqref{vars1}, arbitrary variations of the fields lead to
\[
\bX=\frac{q}m\left(0,\, { -\partial_t \bA -\nabla\Phi+\,}\frac1{|\Psi|^2}\left(\Psi\diamond\frac{\delta h}{\delta \Psi}\right)_{\!\!\bv}\!\times(\nabla\times\bA)\right)-\frac1{m|\Psi|^2}\,\Bbb{J}\!\left(\Psi\diamond\frac{\delta h}{\delta \Psi}\right)
,
\]
along with
\beq\label{HamEM}
\epsilon_0  (\partial_t\bA+\nabla\Phi)  =-\frac{\delta h}{\delta \bE}
\,,\qquad\quad
\epsilon_0\partial_t\bE=\frac{\delta h}{\delta \bA}+\frac{q}{m}\int\!\left(\Psi\diamond\frac{\delta h}{\delta \Psi}\right)_{\!\!\bv}\de^3v\,.
\eeq
Here, the subscript $\bv$ denotes the velocity components of vector-valued quantities in phase-space.
These equations reveal an intricate Hamiltonian structure whose Poisson bracket can be easily obtained by applying the usual relation $\dot{k}=\{\!\{k,h\}\!\}$ upon expanding $\dot{k}=\langle\delta k/\delta\Psi,\partial_t\Psi\rangle+\langle\delta k/\delta\bE,\partial_t\bE\rangle+\langle\delta k/\delta\bB,\partial_t\bB\rangle$ and using the equations of motion for an arbitrary gauge-invariant Hamiltonian  $h$. Indeed,
upon taking the curl of the first equation in \eqref{HamEM} and using $\delta h/\delta\bA=\nabla\times\delta h/\delta\bB$, one obtains the following structure:
\begin{align*}
\{\!\{k,h\}\!\}=&\int\!\frac1{m|\Psi|^2}\!\left(\Psi\diamond\frac{\delta k}{\delta \Psi}\right)\!\cdot\Bbb{J}\!\left(\Psi\diamond\frac{\delta h}{\delta \Psi}\right)
\!\de^3 x\de^3 v+\frac1{\epsilon_0}\!\int\!\left(\frac{\delta k}{\delta \bE}\cdot\nabla\times\frac{\delta h}{\delta \bB}-
\frac{\delta h}{\delta \bE}\cdot\nabla\times\frac{\delta k}{\delta \bB}\right)\!\de^3 x
\\
&\hspace{-1.1cm}+\frac{q}{m\epsilon_0}\!\int\!\left[
\frac{\epsilon_0\bB}{m|\Psi|^2}\cdot\left(\Psi\diamond\frac{\delta k}{\delta \Psi}\right)_{\!\!\bv\!}\times\left(\Psi\diamond\frac{\delta h}{\delta \Psi}\right)_{\!\!\bv\,}
\!+
\left(\Psi\diamond\frac{\delta k}{\delta \Psi}\right)_{\!\!\bv\!}\cdot\frac{\delta h}{\delta \bE}-\left(\Psi\diamond\frac{\delta h}{\delta \Psi}\right)_{\!\!\bv\!}\cdot\frac{\delta k}{\delta \bE}
\right]\!\de^3 x\de^3 v
.
\end{align*}
Then, one can simply evaluate
\[
\frac{\delta h}{\delta \Psi}={mv^2\Psi}
\,,\qquad\quad
\frac{\delta h}{\delta \bE}={\epsilon_0}\bE
\,,\qquad\quad
\frac{\delta h}{\delta \bB}=\mu_0^{-1}\bB
\,,
\]
so that $\PoissonTensor (\Psi\diamond{\delta h}/{\delta \Psi})=(m\bv,0)\norm{\Psi}^2$. 
Introducing the notation ${\widehat{\boldsymbol\Lambda}=-i\hbar(\nabla_{\!\bx},\nabla_{\!\bv})}=({{\boldsymbol\lambda}}_{\bx},{{\boldsymbol\lambda}}_{\bv})$, the KvN equation reads
\[
i\hbar\partial_t\Psi=\bv\cdot{{\boldsymbol\lambda}}_{\bx}\Psi
+\frac{q}{m}\left(\bE+\bv\times\bB\right)\cdot{{\boldsymbol\lambda}}_{\bv}\Psi
\,.
\]
which is accompanied by Faraday's and Amp\'ere's law
\[
\partial_t\bB=-\nabla\times\bE
\,,\qquad\quad
\epsilon_0\partial_t\bE=\mu_0^{-1}\nabla\times\bB-q\int\!|\Psi|^2\bv\,\de^3v
\,.
\]
Here, the first equation is obtained by taking the curl of the first equation in \eqref{HamEM}. 

\section{Koopman-van Hove theory \label{sec:KvH_theory}}
As we saw in previous sections, the KvN phase remains constant along the  phase-space Lagrangian trajectories;  see the second equation in \eqref{DSeqs}. 
Indeed, in the KvN construction the phase is entirely irrelevant and plays the role of a gauge freedom in the relation $f=|\Psi|^2$. 
However, in classical mechanics one usually relates the classical phase to the Lagrangian function. 
Specifically, in Hamilton-Jacobi theory the phase is a function in configuration space and this function is given by the classical action integral. 
We observe that this is different from the Koopman phase, which instead is defined on phase-space. 
As we shall see, the Koopman-van Hove construction 
 combines the theory of Koopman wavefunctions with 
 Feynman's prescription of a phase expressed in terms of the Lagrangian.
Again, to keep the notation as simple as possible, the phase space will be two-dimensional in most of our discussion. 
The extension to six dimensions is straightforward and will be used when coupling the KvH equation to the electromagnetic fields.

\subsection{The Koopman-van Hove equation \label{sec:KvH_equation}}
 
An alternative theory of classical mechanics based on Koopman wavefunctions goes back to van Hove's thesis \cite{VanHove}, where canonical transformations were extended to include phase factors. 
As later shown by Kostant \cite{Ko1970}, in this setting the phase function is again identified with the action integral, which is now defined in terms of the phase-space Lagrangian. 
In this setting, the KvN equation  \eqref{KvN-eq} becomes the \emph{\color{black}Koopman-van Hove equation}
\beq\label{KvH-eq}
i\hbar\partial_t\Psi=\widehat{L}_H\Psi-\mathscr{L}\Psi\,,
\qquad\quad\text{with}\qquad\quad
\mathscr{L}=
p\partial_p H-H.
\eeq
Here $\widehat{L}_H=\partial_xH{\lambda}_p-\partial_pH{\lambda}_x$ while $\mathscr{L}$ is the expression of the phase-space Lagrangian, which now identifies a phase term.  The information contained in this equation may be unfolded by applying the Madelung transform $\Psi=\sqrt{D}e^{iS/\hbar}$, which leads to
\beq\label{Madelung-KvH}
\partial_t D+\{D,H\}=0
\,,\qquad\quad
\partial_t S+\{S,H\}=\mathscr{L}
.
\eeq
The second equation can be formally solved in terms of the Lagrangian trajectories as \cite{GBTr20,GBTr21a}
\beq\label{KvHphase1}
S(\bz,t)=\int_{0}^t\mathscr{L}(\boldsymbol{\eta}(\bz,\tau-t))\,\de\tau+ S_0(\boldsymbol{\eta}^{-1}(\bz,t))
\,,
\eeq
thereby showing how the phase evolution emerges from the integral of the Lagrangian, in analogy to Feynman's  path-integral formulation of quantum mechanics. {\color{black}For a discussion of the relation between equation \eqref{KvHphase1} and the Hamilton-Jacobi equation, see de Gosson's work in \cite{deGo04}.}

\begin{remark}[\color{black}Relation to hybrid quantum-classical dynamics]
Notice that{\color{black}, for Hamiltonians of the type $H=T+V$,} enforcing $\partial_p\Psi=0$ and replacing $p\to-i\hbar\partial_x$ takes \eqref{KvH-eq} into the quantum Schr\"odinger equation, thereby justifying the early name \emph{prequantum Schr\"odinger equation} \cite{Ko1970}. Indeed, equation \eqref{KvH-eq} first appeared within the context of prequantization theory \cite{Kirillov,VanHove} and it has remained pretty unknown over the decades. Recently, it was recognized how this equation may actually lead to a consistent theory of quantum-classical coupling \cite{BoGBTr19,GBTr20,GBTr21b}, where the phase plays a crucial role. Partly inspired by Kirillov \cite{Kirillov}, the authors of \cite{BoGBTr19,GBTr20} called equation \eqref{KvH-eq}  `Koopman-van Hove equation' in recognition of the very first contributions from Koopman and van Hove.
\end{remark}

\subsection{Geometric setting \label{sec:vanHove_transformations} }
While canonical transformations are enough to characterize the evolution of Koopman wavefunctions in KvN theory, the presence of the phase in the KvH formalism requires extending the KvN picture. A more detailed summary of the geometric setting of KvH theory is found in \cite{Faure,GBTr20,GBTr21a}.  First, one introduces the $U(1)-$bundle $\Bbb{R}^{2}\times U(1)$, where $\Bbb{R}^{2}$ is the Euclidean two-dimensional phase-space and $U(1)$ is the group of complex phase factors. Gauge transformations are identified, as usual, with local phase factors so that the KvH wavefunction evolves according to compositions of gauge transformations and canonical transformations, that is
\beq\label{KvHevol}
\Psi(\bz,t)=\left.e^{-i\varphi(\bz_0,t)/\hbar\,}\Psi_0(\bz_0)\right|_{\bz_0=\boldsymbol\eta^{-1}(\bz,t)}
.
\eeq
Notice that, unlike \eqref{HDevol}, here we are restricting the Lagrangian trajectory to identify a canonical transformation at all times.
The geometric characterization  of the phase factor in \eqref{KvHevol} needs further discussion. Specifically, the relation between the phase factor and the phase-space Lagrangian emerges as follows. Upon defining the gauge connection $\boldsymbol{\cal A}=p \de x$, it is well known \cite{MaRa} that $\de(\eta^*\boldsymbol{\cal A}-\boldsymbol{\cal A})=0$, where $\de$ is the exterior differential and $\eta^*\boldsymbol{\cal A}(\bz):={\cal A}_\ell(\boldsymbol\eta(\bz))\nabla\eta^\ell(\bz)$ is the standard  pullback of the connection one-form $\cal A$ by the canonical transformation $\eta$. Then, the  phase factor in \eqref{KvHevol} is defined via
\beq\label{KvHphase2}
\de\varphi:=\boldsymbol{\cal A}-\eta^*\boldsymbol{\cal A}
\,.
\eeq
Indeed, upon using the  Lie derivative theorem $\de(\eta^*\boldsymbol{\cal A})/\de t=\eta^*\pounds_{\bX_H}\boldsymbol{\cal A}$, we notice that Cartan's magic formula takes the time derivative of \eqref{KvHphase2} into the form
\[
\partial_t\de\varphi=-\eta^*\pounds_{\bX_H}\boldsymbol{\cal A}=-\eta^*\de({\bX_H}\cdot\boldsymbol{\cal A}-H)= -\de(\eta^*\Lag)
\,,
\qquad\ \text{ with }\qquad\ 
\Lag:=\bp\cdot \partial_\bp  H-H
\,,
\]
so that $\partial_t\varphi(\bz_0,t)=\omega(t)-\mathscr{L}(\boldsymbol\eta(\bz_0,t))$. Then, up to a time-dependent frequency $\omega(t)$, equation \eqref{KvHphase1} follows from the defining relation $S(\bz,t):=(S_0(\bz_0)-\varphi(\bz_0,t)|_{\bz_0=\boldsymbol\eta^{-1}(\bz,t)}$, where $\Psi_0=\sqrt{D_0}e^{iS_0/\hbar}$.

The evolution law \eqref{KvHevol} together with the definition \eqref{KvHphase2} represents the KvH analogue of \eqref{HDevol} from KvN theory. The  propagator $\Psi_0(\bz)\mapsto\Psi(\bz,t)$ was first devised by van-Hove \cite{VanHove} and identifies a unitary transformation called a \emph{van Hove transformation} in \cite{GBTr20,GBTr21a}. 
Given a canonical transformation $\eta$, the van Hove transformation  reads
\beq
U_\eta\Psi(\bz):=e^{-i\varphi(\bz_0)/\hbar\,}\Psi_0(\bz_0)|_{\bz_0=\boldsymbol\eta^{-1}(\bz)}
.
\label{VHTransf}
\eeq
 Without going much into the details, here we shall simply point out that van Hove transformations possess a group structure whose Lie algebra is identified with the space of Hamiltonian functions endowed with the Lie bracket given by the canonical Poisson bracket. In turn, the self-adjoint operator 
\beq\label{prequantop}
\widehat{\cal L}_{H}=\widehat{L}_H+H-p \partial_p H = \KvNOp -\Lag
\eeq
in \eqref{KvH-eq} identifies the infinitesimal generator $-i\hbar^{-1}\widehat{\cal L}_{H}$ of van Hove transformations. Since KvH theory  emerged historically in prequantization theory, here we shall keep the standard nomenclature by calling \eqref{prequantop} the \emph{prequantum operator}.

\subsection{The phase-space density\label{sec:KvHdensity}}
So far, nothing has been said about the relation between KvH wavefunctions and the classical phase-space density. In principle, one could insist on following Koopman's original prescription $f=|\Psi|^2$. However, as we will see shortly, this step poses  questions similar to those arising in Section  \ref{sec:KvN-canonical}.
At present, the only Hamiltonian structure available for the KvH equation \eqref{KvH-eq} is given by the canonical bracket \eqref{CanPB}, which is accompanied by the Hamiltonian functional
\beq\label{KvHHam}
h_{\text{\tiny KvH}}(\Psi)=\int\!\Psi^*\widehat{\cal L}_H\Psi\,\de x\de p=\int\!H\big(|\Psi|^2+\partial_p(p|\Psi|^2)+\hbar\operatorname{Im}\{\Psi^*,\Psi\}\big)\,\de x\de p
\,,
\eeq
where we notice that $\partial_p(p|\Psi|^2)=\operatorname{div}(\Bbb{J}\boldsymbol{\cal A}|\Psi|^2)$.
\addedtext{
}
If we follow Koopman's original prescription of using $f=|\Psi|^2$ for the phase-space density, then 
the Hamiltonian functional does not generally coincide with the total energy of the system.
As shown in \cite{BoGBTr19,GBTr20,GBTr21a}, the expression in parenthesis in \eqref{KvHHam} identifies the momentum map for the (left) unitary representation \eqref{VHTransf} of van Hove transformations on the Koopman Hilbert space ${\cal H}=L^2(\Bbb{R}^2)$. Consequently, {\color{black}by the arguments in Remark \ref{Collectivization}}, the identification
\beq\label{rhomomap}
f=|\Psi|^2+ \partial_p (p |\Psi|^2 )+\hbar\operatorname{Im}\{\Psi^*,\Psi\}
\eeq
leads to the usual Liouville equation $\partial_t f=\{f,H\}$. While it may be objected that $f$ is not positive definite, the flow of $f$ preserves the sign of the initial condition thereby eliminating this apparent problem.

Nevertheless, the identification \eqref{rhomomap} represents a change of perspective from the conventional Koopman {\color{black}prescription $f=|\Psi|^2$} in that the KvH phase $S$ enters the expression of the phase-space density. However, while {\color{black} the $f$ given in \eqref{rhomomap}} comprises the entire physical information, we emphasize that in the present formalism the single terms in \eqref{rhomomap} do not possess any physical meaning despite the fact that {\color{black}both the first and the sum of the last two}  obey the Liouville equation. In particular, this representation   of the phase-space density identifies an alternative Clebsch representation extending the usual case given by the last term in \eqref{rhomomap}; see Section \ref{sec:KvN-canonical}. Further discussions are found in \cite{GBTr20,GBTr21a}. A point of relevance for later purpose is that the momentum map $f(\Psi)$ in \eqref{rhomomap} is covariant (or \emph{equivariant}) with respect to canonical transformations. 
Indeed, 
using the notation of Eq.  \eqref{VHTransf}, we have
\beq\label{equivariance}
f(U_\eta\Psi)
=f\circ{\boldsymbol\eta}^{-1}
\,.
\eeq
The details of this property can be found in eg. \cite{GBTr20}.

\begin{remark}[\color{black}Constraints and phase singularities] {\color{black}Notice that here we can follow the same arguments as in Section  \ref{sec:KvN-canonical} in order to 
 enforce the relation $\pdf=\norm{\WaveFunc}^2$ as a specific constraint. Indeed, as hinted in \cite{GBTr20}, one may be tempted to write $\Psi=\sqrt{D}e^{iS/\hbar}$ and choose the phase $S$ so that
\[
\partial_p (p |\Psi|^2 )+\hbar\operatorname{Im}\{\Psi^*,\Psi\}=
\operatorname{div}\left(D \Bbb{J}(\nabla S-\boldsymbol{\cal A})\right)=0.
\]
As} proven in Ref. \cite{Joseph20}, if this relation holds for the initial condition, then it will hold for all time.
For an arbitrary $D$, this can only be true if $\nabla S=\boldsymbol{\cal A}$, which simultaneously requires $\partial S/\partial p =0$ and $\partial S/\partial{x}  = p$.
For general nonlinear systems, this solution requires $S$ to have special types of coordinate singularities when the Hamiltonian flow has O-points and X-points \cite{Joseph20}.
For an integrable system where action-angle coordinates $\{J,\theta\}$ can be constructed, the Hamiltonian is a function of the action variables $J$ alone, $H(J)$. 
In this case, the solutions to Hamilton's equations of motion simplify to have the form $\theta=\theta_0+\omega(J) t$, where $\omega=\partial H /\partial J$.
The general solution to the Hamilton-Jacobi equation has the form $S=S_0+J \theta - H(J) t$, where $S_0$ is an arbitrary function of constants of the motion.
This can also be written as $S=S_0 +J\theta_0 + (J \omega-H) t$; i.e. as the sum of an arbitrary constant of the motion and the Lagrangian multiplied by the time.
For quadratic Hamiltonians, such as the harmonic oscillator, the general solution clearly reduces to $S=S_0$, an arbitrary function of constants of the motion.
\end{remark}

{\color{black}Before concluding this Section, we notice}  the expression{\color{red}s} of the first two moments
\[
\int f\,\de p=
\int|\Psi|^2+\hbar\operatorname{Im}\int\{\Psi^*,\Psi\}\,\de p
\,,\qquad\qquad
\int pf\,\de p=
\hbar\operatorname{Im}\int p\{\Psi^*,\Psi\}\,\de p
\,,
\]
which will play a crucial role in the coupling to electromagnetic fields, as shown in the next section. In more generality, the relation
\[
\int\! p^{m} f\,\de p=
\int \!p^m\Big((1-m)|\Psi|^2+\hbar\operatorname{Im}\{\Psi^*,\Psi\}\Big)\de p
\]
provides an alternative representation of Vlasov moments in terms of canonical variables.


\subsection{KvH-Maxwell system and its
\label{sec:KvH-Maxwell}
Hamiltonian structure}

In this section, we apply the KvH formalism to the Vlasov kinetic theory of magnetized plasmas. In particular, we are interested in the Hamiltonian structure for the system comprising the electromagnetic component as well as the KvH wavefunction $\Psi(\bx,\bv)$ expressed in terms of the velocity variable $\bv=(\bp-q\bA)/m$.
As customary in the geometric approach to the Maxwell-Vlasov system, we start in terms of canonical variables and write the action principle for an arbitrary gauge
\beq\label{CanVPKvHMax}
\delta\int^{t_1}_{t_0}\Big(\big\langle\Psi,i\hbar\partial_t\Psi-q\widehat{\cal L}_\Phi\Psi\big\rangle-\epsilon_0\big\langle\bE,\partial_t\bA+\nabla\Phi\big\rangle-h(\Psi,\bA,\bE)
\Big)\,\de t=0
\,,
\eeq
where 
\beq\label{canHam}
h(\Psi,\bA,\bE)={h_{\text{\tiny KvH}}(\Psi) }
+h_\text{\tiny Max}(\bE,\bA)
\qquad\quad\text{and}\qquad\quad
H=\frac1{2m}\left|\bp-q\bA\right|^2
\,.
\eeq
Here, $h_\text{\tiny Max}(\bE,\bA)$ is the standard Maxwell Hamiltonian \eqref{MaxHam} while $h_{\text{\tiny KvH}}(\Psi)=\langle\Psi,\widehat{\cal L}_H \Psi\rangle$ involves the prequantum operator  \eqref{prequantop}, so that $\langle\Psi,\widehat{\cal L}_H \Psi\rangle=\langle f, H\rangle$ with $f$ given in \eqref{rhomomap}. Similarly $\langle\Psi,\widehat{\cal L}_\Phi\Psi\rangle=\int 
\!f\Phi\,\de^3 x\de^3 p$ and $\Phi$ plays again the role of a Lagrange multiplier enforcing Gauss' law as in \eqref{KMVP}. {\color{black}Notice that here one may choose to include the $\Phi-$terms in the Hamiltonian, which would then become a Routhian. However, in order to obtain an explicit Hamiltonian structure comprising a Poisson bracket, it is customary to fix the Hamiltonian gauge $\Phi=0$ in \eqref{CanVPKvHMax} so that the resulting Hamilton's equations yield}
\beq\label{canPBKvHMax}
\{\!\{k,h\}\!\}(\Psi,\bA,\bE)=\frac1{2\hbar}\operatorname{Im}\left\langle\frac{\delta h}{\delta \Psi}\bigg|\frac{\delta k}{\delta \Psi}\right\rangle-\frac1{\epsilon_0}\left\langle\frac{\delta k}{\delta \bA},\frac{\delta h}{\delta \bE}\right\rangle+\frac1{\epsilon_0}\left\langle\frac{\delta k}{\delta \bE},\frac{\delta h}{\delta \bA}\right\rangle
.
\eeq
We recall that the angle brackets denote the standard $L^2-$pairing. Then, the KvH equation $i\hbar\partial_t\Psi=\widehat{\cal L}_H \Psi$ is accompanied by $\partial_t\bA=-\bE$ and Amp\`ere's law in the form
\begin{align*}
\epsilon_0\partial_t\bE=&\ \mu_0^{-1}\nabla\times\bB-\frac{q}m
\int(\bp-q\bA)f\de^3 p
\\
=&\ 
\mu_0^{-1}\nabla\times\bB-\frac{q\hbar}m\operatorname{Im}
\int\!\bp \{\Psi,\Psi^*\}\,\de^3 p
+
\frac{q^2}m\bA\int\!\left(|\Psi|^2+\hbar\operatorname{Im}\{\Psi,\Psi^*\}\right)\,\de^3 p
\end{align*}

An immediate way of obtaining the Hamiltonian structure in terms of $\Psi(\bx,\bv)$ is given by a direct change of coordinates.   
Here, we notice the convenient abuse  in denoting the Koopman wavefunction of both canonical and noncanonical coordinates by the same symbol $\Psi$. 
The coordinate change $(\bx,\bp)\to(\bx,\bv)=(\bx,(\bp-q\bA)/m)$ corresponds to the replacement
\[
\frac{\delta}{\delta \bA}\to\frac{\delta}{\delta \bA}+\frac{q}m\operatorname{Re}\int\!\de^3v\,\nabla_{\!\bv}\Psi^*\frac{\delta}{\delta \Psi}
\,,
\]
which in turn transforms \eqref{canPBKvHMax} into
\begin{align}\nonumber
\{\!\{k,h\}\!\}(\Psi,\bA,\bE)=&\ \frac1{2\hbar}\operatorname{Im}\left\langle\frac{\delta h}{\delta \Psi}\bigg|\frac{\delta k}{\delta \Psi}\right\rangle-\frac1{\epsilon_0}\left\langle\frac{\delta k}{\delta \bA},\frac{\delta h}{\delta \bE}\right\rangle+\frac1{\epsilon_0}\left\langle\frac{\delta k}{\delta \bE},\frac{\delta h}{\delta \bA}\right\rangle
\\
&\ -\frac{q}{m\epsilon_0}
\left\langle\frac{\delta k}{\delta \Psi},
\frac{\delta h}{\delta \bE}\cdot\nabla_{\!\bv}\Psi\right\rangle+\frac{q}{m\epsilon_0}
\left\langle\frac{\delta h}{\delta \Psi},
\frac{\delta k}{\delta \bE}\cdot\nabla_{\!\bv}\Psi\right\rangle
.
\label{PBKvHNC}
\end{align}
Notice that, under the same change of variables $(\bx,\bp)\to(\bx,\bv)$, the prequantum operator \eqref{prequantop} becomes
\beq 
\widehat{\cal L}_H=\frac{1}m
 \widehat{L}_H  +i \frac{q\hbar}{m^2}\bB\cdot\frac{\partial H}{\partial\bv} \times\frac{\partial}{\partial\bv}+H-\Big(\bv+\frac{q}m\bA\Big)\cdot\frac{\partial H}{\partial\bv}\,.
\label{NCpreq}
\eeq
Unless otherwise specified,  in the remainder of this section we shall restrict to the case $H={m}v^2/2$, as in \eqref{canHam}.
Also, the expression \eqref{rhomomap} of the phase-space density becomes 
\beq
f=
|\Psi|^2+\nabla_{\!\bv}\cdot\Big(\Big(\bv+\frac{q}m\bA\Big)|\Psi|^2\Big)+\frac{\hbar}{m}\operatorname{Im}\Big(\{\Psi^*,\Psi\} + \frac{q}m \bB\cdot\nabla_{\!\bv}\Psi^*\times\nabla_{\!\bv}\Psi\Big)
\,.
\label{NCmomap}
\eeq
In conclusion, upon writing $({{\boldsymbol\lambda}}_{\bx},{{\boldsymbol\lambda}}_{\bv})=-i\hbar(\nabla_{\!\bx},\nabla_{\!\bv})$, the KvH equation $i\hbar(\partial_t\Psi
+qm^{-1}\bE\cdot\nabla_{\!\bv})\Psi=\widehat{\cal L}_{H}\Psi$
reads 
\beq
i\hbar\partial_t\Psi=
\bv\cdot\widehat{\boldsymbol{\lambda}}_\bx\Psi+\frac{q}{m}\left(\bE+\bv \times\bB\right)\cdot\widehat{\boldsymbol{\lambda}}_{\bv}\Psi -\Big(\frac{m}2v^2+q\bv\cdot\bA\Big)\Psi
\,,
\label{NCKvH}
\eeq
which is accompanied by $
 \partial_t\bA=-\bE$ and Amp\`ere's law in the form
\beq\label{KvHMaxHam}
\epsilon_0\frac{\partial\bE}{\partial t}=\frac{\delta h}{\delta\bA}+\frac{q}m\operatorname{Re}\int\!(\nabla_{\!\bv}\Psi)^*\frac{\delta h}{\delta\Psi}\,\de^3v
\qquad\text{with}\qquad
h=h_\text{\tiny Max}(\bE,\bA) + h_{\text{\tiny KvH}}(\Psi)
\,.
\eeq
More explicitly, one has   $\epsilon_0\partial_t\bE-\mu_0^{-1}\nabla\times\bB=-\bJ$ where
\begin{align*}
- \bJ &  =\frac{\delta h_{\text{\tiny KvH}}}{\delta \bA}  +
 \frac{q}m\operatorname{Re}\int\!(\nabla_{\!\bv}\Psi)^*\frac{\delta h_{\text{\tiny KvH}} }{\delta\Psi}\,\de^3v 
\\
& =  
\
\frac{q}m
\int\!|\Psi|^2\nabla_{\!\bv}H\,\de^3 v
+\frac{q\hbar}{m^2}
\nabla\times\int\!H\operatorname{Im}(\nabla_{\!\bv}\Psi^*\times\nabla_{\!\bv}\Psi)\,\de^3 v  
+2
\frac{q}m\operatorname{Re}\int\!(\nabla_{\!\bv}\Psi)^*\widehat{\cal L}_H \Psi\,\de^3v  
\\
&= \
 q\int \!H\nabla_\bv f\,\de^3v
\\
&= \ 
-q\int\!\bv f\,\de^3 v
\,.
\end{align*}
In order to derive these relations, one makes repeated use of integration by parts in combination with \eqref{NCpreq}, \eqref{NCmomap}, and
\begin{align*}
\operatorname{Im}\int\!
\{H, \Psi\}\nabla_{\!\bv}\Psi^*\,\de^3v
=&\, \operatorname{Im}
\int\!
H\{ \nabla_{\!\bv}\Psi^*,\Psi\}\,\de^3v-
\operatorname{Im}
\int\!
\{ \Psi,H\nabla_{\!\bv}\Psi^*\}\,\de^3v
\\
=& 
\, \operatorname{Im}
\int\!
H\{ \nabla_{\!\bv}\Psi^*,\Psi\}\,\de^3v
-\frac12 \nabla\times\int\!H\operatorname{Im}(\nabla_{\!\bv}\Psi^*\times\nabla_{\!\bv}\Psi)\,\de^3 v
\,.
\end{align*}
Here, the last equality follows from $\operatorname{Im}
\int
\{ \Psi,H\nabla_{\!\bv}\Psi^*\}\,\de^3v=-\operatorname{div}\int H\operatorname{Im}(\nabla_{\!\bv}\Psi\otimes\nabla_{\!\bv}\Psi^*)\,\de^3v$ by using $\operatorname{Im}\int\!\Psi^*\nabla_{\bv}\nabla_{\bv}\Psi\,\de^3v=0$ and standard vector algebra. 
Notice that we have the relation
\[  
\hbar\frac{\partial}{\partial t}\!\left(\operatorname{Im}\int\!
 \Psi^* \nabla_{\bv}\Psi 
\,\de^3 v\right)
=-2\operatorname{Re}\int\!(\nabla_{\!\bv}\Psi)^*\widehat{\cal L}_H \Psi\,\de^3v
\,,
\]
which follows directly from the KvH equation $i\hbar(\partial_t\Psi
+qm^{-1} \bE\cdot\nabla_{\!\bv})\Psi=\widehat{\cal L}_{H}\Psi$ for an arbitrary function $H$.

\subsection{Gauge invariance and charge conservation\label{sec:gaugetransf}}
So far, nothing has been said about the role of gauge transformations. In this section, we will show how Gauss's law 
\beq\label{GLaw}
\operatorname{div}(\epsilon_0\bE)=q \int \!f\,\de^3p =q\int|\Psi|^2\,\de^3 p+q\hbar\operatorname{Im}\int\{\Psi,\Psi^*\}\,\de^3 p
\eeq
arises as usual from gauge invariance. 
It is well known that Gauss law arises from the symmetry of the simultaneous action of gauge transformations over both vector potentials and phase-space quantities. In order to avoid unnecessary difficulties, as explained in \cite{MaWe1}, it is convenient to study the properties of gauge transformations in terms of the canonical coordinates $(\bx,\bp)$. In particular, besides the standard gauge transformation on the electromagnetic quantities  $(\bA,\bE)\mapsto(\bA+\nabla\chi,\bE)$, phase-space coordinates undergo  momentum translations of the type $(\bx,\bp)\mapsto(\bx,\bp+q\nabla\chi)$. In turn, this produces an action of gauge transformations on the phase-space density that is $f(\bx,\bp)=f(\bx,\bp-q\nabla\chi)$. The latter is a type of canonical transformation, which will be the starting point of our discussion.

In order to examine the role of gauge transformations, we have to construct an action of momentum translations on the space of KvH wavefunctions. If we were dealing with KvN theory, this would be simply given by $\Psi_\text{\tiny KvN}(\bx,\bp)\mapsto\Psi_\text{\tiny KvN}(\bx,\bp-q\nabla\chi)$. However, KvH wavefunctions also carry a phase factor which we now turn to. If $\boldsymbol{\eta}(\bx,\bp)=(\bx,\bp+q\nabla\chi)$ identifies a momentum translation, equation \eqref{KvHphase2} yields
\[
\nabla\varphi=\bp-(\bp+q\nabla\chi)=-q\nabla\chi
\]
so that $\varphi=-q\chi$ and in this case the phase-space function $\varphi$ depends only on the spatial coordinates $\bx$. Then, since $\chi(\boldsymbol{\eta}^{-1}(\bz))=\chi(\bx)$, we are led to the following unitary action $\Psi_\text{\tiny KvH}\mapsto U_{\eta}\Psi_\text{\tiny KvH}$ of momentum translations on KvH wavefunctions:
\[
\Psi_\text{\tiny KvH}(\bx,\bp)\mapsto e^{iq\chi/\hbar\,} \Psi_\text{\tiny KvH}(\bx,\bp-q\nabla\chi)
=U_{\eta}\Psi_\text{\tiny KvH}(\bx,\bp)
\,.
\]
Upon dropping the subscript `KvH', we will now show that the Hamiltonian \eqref{canHam} is invariant under the gauge transformation
\[
\left(\Psi(\bx,\bp),\bA,\bE\right)\mapsto\left(e^{iq\chi/\hbar\,} \Psi(\bx,\bp-q\nabla\chi),\bA+\nabla\chi,\bE\right)
\,.
\]
Evidently, $h_\text{\tiny Max}$ in \eqref{MaxHam} is manifestly gauge-invariant and thus here we consider the first term $\langle\Psi,\widehat{\cal L}_H \Psi\rangle$. The gauge invariance of this functional is an immediate consequence of the fact that the phase-space density  \eqref{rhomomap} identifies an equivariant momentum map. Indeed, since momentum translations are canonical transformations, we can use \eqref{equivariance} to write
\begin{align*}
\langle U_\eta\Psi,\widehat{\cal L}_H U_\eta\Psi\rangle
&\ =\langle f\circ\boldsymbol{\eta}^{-1},H\rangle
\\
&\ =\int\! f(\bx,\bp)\,H(\bx,\bp+q\nabla\chi)\,\de^3 x\,\de^3 p
\\
&\ 
=
\frac1{2m}\int\!\left|\bp+q\nabla\chi-q\bA\right|^2\de^3 x\,\de^3 p
\,.
\end{align*}
Then, the transformation $\bA\mapsto\bA+\nabla\chi$ leads to overall gauge invariance of $\langle\Psi,\widehat{\cal L}_H \Psi\rangle$.

At this point, we have characterized the gauge transformations that leave the Hamiltonian $h(\Psi,\bA,\bE)=\langle\Psi,\widehat{\cal L}_H \Psi\rangle+h_\text{\tiny Max}(\bE,\bA)
$ invariant and we are ready to present the associated conserved quantity. Here, we shall proceed once again by exploiting momentum maps: since the action of gauge transformations on $(\Psi,\bA,\bE)$ leaves the Hamiltonian invariant, the momentum map associated to this action is conserved by the dynamics. This momentum map must satisfy the defining relation \eqref{Momapdef}. In this case, the Poisson bracket is given by \eqref{canPBKvHMax} and we have $M=L^2(T^*Q)\times T^*\Omega^1(Q)$, where  $\Omega^1(Q)$ denotes the space of differential one-forms on $Q$ and $Q=\Bbb{R}^3$. As discussed in \cite{MaWe1}, the Lie algebra of gauge transformations is identified with smooth scalar functions $\xi(\bx)$ on $Q$, so that their infinitesimal generator reads
\[
\xi_M(\Psi,\bA,\bE)=\big(-iq\hbar^{-1}\widehat{\cal L}_{\xi}\Psi,\nabla\xi,0\big)
\,.
\]
Then, upon writing the momentum map 
\[
{\cal J}(\Psi,\bA,\bE)   
=\operatorname{div}(\epsilon_0\bE)-q\int|\Psi|^2\,\de^3 p-q\hbar\operatorname{Im}\int\{\Psi,\Psi^*\}\,\de^3 p=  \operatorname{div}(\epsilon_0\bE)-q\int\! f\,\de^3 p
\] 
and computing
$
\langle{\cal J}(\Psi,\bA,\bE),\xi\rangle=-\epsilon_0\langle\bE,\nabla\xi\rangle-q\langle\Psi,\widehat{\cal L}_{\xi}\Psi\rangle
$, one indeed verifies \eqref{Momapdef}. Thus, Gauss Law \eqref{GLaw} emerges as the zero-level set of a conserved momentum map associated to the action of gauge transformations.

We conclude our discussion by noticing that, while the Hamiltonian \eqref{canHam} is gauge invariant, its dependence on $\bA$ cannot be generally expressed only in terms of the magnetic field $\bB=\nabla\times\bA$. While this is precisely what happens also in the standard Hamiltonian treatment of the Maxwell-Vlasov system, here we observe that this feature persists after changing to noncanonical coordinates and this is due to the presence of the Lagrangian function in the prequantum operator \eqref{NCpreq}. As proposed in \cite{MaWe1}, one can still write the Hamiltonian $h(\Psi,\bA,\bE)$ in \eqref{KvHMaxHam} in terms of $(\Psi,\bB,\bE)$ at the expenses of fixing a convenient gauge such as the Coulomb gauge or the Poincar\'e gauge. For example, the Coulomb gauge yields $\bA=\nabla\times(\Delta^{-1}\bB)$, which can then be replaced in the expression \eqref{NCpreq} of the prequantum operator. Then, the Poisson bracket \eqref{PBKvHNC} also changes according to the familiar chain rule relation $\delta/\delta\bA=\nabla\times\delta/\delta\bB$.

\section{Discussion}
 
The KvN and KvH equations represent two valid approaches to developing a Hilbert space formulation of classical mechanics on phase space. 
Both approaches lead to a generalized Clebsch representation for the PDF $\pdf$.
A specific choice of the complex phase factor allows this Clebsch representation for $\pdf$ to become equal to the Koopman prescription $\norm{\WaveFunc}^2$.
However, this choice requires the phase factor to become {\color{black}singular}.

In both formulations, the complex phase factor {\color{black}is generally involved in reproducing} the classical dynamics. 
In fact, for the KvN formulation, the additional phase degree of freedom is formally required for obtaining a canonical variational formulation.
This canonical KvN-Maxwell formulation parallels the development of the canonical KvH-Maxwell formulation, and can be obtained from the results of Sec. \ref{sec:KvH_theory} by simply eliminating the Lagrangian from the definition of $\widehat{\cal L}_{H}$. In fact, this canonical KvN-Maxwell formulation has been treated in \cite{Neiss} in the electrostatic limit. Alternatively,
 in order to eliminate the need to include the phase in KvN dynamics,  a noncanonical Poisson bracket was determined that reduces to the standard Vlasov  bracket  for functionals that depend only on $|\Psi|^2$.
In this case, the phase becomes completely irrelevant and can be set to be identically zero.

The ``pre-quantum''  KvH formulation begins to bridge the gap between the classical and quantum mechanical dynamics by providing a physically motivated prescription for the evolution of the phase factor that agrees with the semiclassical $\hbar\rightarrow0$  limit.
Thus, the prequantum KvH equation can begin to describe some of the important physical consequences of coupling a {\color{black}classical} system to a truly quantum system \cite{BoGBTr19,GBTr20,GBTr21b}.
In contrast, the KvN formulation, with trivial phase dynamics, is perhaps better considered to correspond to the diagonal part of the density matrix.

In dealing with both canonical and noncanonical structures, some comments on their numerical aspects are also in place. Indeed, 
both  symplectic numerical integrators and quantum simulation algorithms are well understood for canonical Hamiltonian systems, but not for noncanonical systems with an arbitrary Poisson bracket.  While this makes the canonical KvN and KvH formulations more amenable to the development of numerical integration techniques that preserve conservation laws, it also motivates future research on developing numerical methods that target the new noncanonical KvN formulation derived here.

Both approaches can be used to develop a ``quantum'' representation of the classical Liouville equation and both can be simulated on a quantum computer.
However, once the Koopman equation is coupled to Maxwell's equations, one obtains a coupled system of nonlinear partial differential equations. 
It is only possible to efficiently simulate these equations using a quantum computer if they are embedded within a unitary linear system of equations.
This can be done by simulating the classical statistical probability density, ${\mathscr F}(\Psi,\bA,\bE;t,\bx)$, for the fields at every point in space-time.
As described in Ref. \cite{Joseph20}, the Liouville equation  for  ${\mathscr F}$ can  be simulated efficiently on a quantum computer.
Understanding the complexity of quantum simulation for each of these Koopman-Maxwell formulations is an important topic for future research.

{\color{black}
\paragraph{Acknowledgments.} We wish to thank our colleagues Denys Bondar, Joshua Burby, Fran\c{c}ois Gay-Balmaz, John Finn, Michael Kraus, Omar Maj, and Philip Morrison for several interesting discussions on this and related topics. The work of CT is partially supported by the Royal Society, UK. The work of IJ was performed under the auspices of the U.S. Department of Energy (DOE) by Lawrence Livermore National Laboratory (LLNL) under Contract DE-AC52- 07NA27344. IJ was supported by the DOE Office of Fusion Energy Sciences “Quantum Leap for Fusion Energy Sciences” project FWP-SCW1680 and by LLNL Laboratory Directed Research and Development Project 19-FS-072.
}


\begin{thebibliography}{99}



\bibitem{Alanson92}
{Alanson, T. {\it A ``quantal" Hilbert space formulation for nonlinear dynamical systems in terms of probability amplitudes}, Phys. Lett. A 163 (1992) 41-45.}

\bibitem{BatesWeinstein}
Bates, S.; Weinstein, A. {Lectures on the Geometry of Quantization}. Berkeley Mathematics Lecture Notes. 8, AMS, Providence, 1997.

 
\bibitem{Berry}
Berry, M.V.; {\it True quantum chaos? An instructive example}. In ``New Trends in Nuclear Collective Dynamics''. Edited by Y. Abe, H. Horiuchi, and K. Matsuyanagi. Springer-Verlag Berlin Heidelberg. 1992

\bibitem{Bondar}
Bondar, D.I.; Cabrera, R.; Lompay, R.R.; Ivanov, M.Yu.; Rabitz, H.A. {\it Operational dynamic modeling transcending quantum and classical mechanics}. {Phys. Rev. Lett.} 109 (2012), 190403 

\bibitem{BoGBTr19}
Bondar, D.I.; Gay-Balmaz, F.; Tronci, C. {\it  Koopman wavefunctions and classical-quantum correlation dynamics}. Proc. R. Soc. A 475 (2019), n. 2229, 20180879

%

\bibitem{boucher}
Boucher, W.; Traschen, J. {\it Semiclassical physics and quantum fluctuations}. {Phys. Rev. D} 37 (1988), 3522-3532

\bibitem{Mezic}
Budi\v{s}i\'{c}, N.; Mohr, R.; Mezi\'{c}, I. {\it Applied Koopmanism}. Chaos 22 (2012), 047510.

\bibitem{Cendra}
Cendra, H.; Holm, D.D.; Hoyle, M.J.W.; Marsden, J.E. {\it  The Maxwell-Vlasov equations in Euler-Poincar\'e form}. J. math. Phys. 39 (1998), n. 6, 3138-3157 

\bibitem{ChMa}
Chernoff, P.R.; Marsden, J.E. {\it Some remarks on Hamiltonian systems and quantum mechanics}. 
Univ. Western Ontario Ser. Philos. Sci. 6c (1976), 35-53

\bibitem{Chirikov88}
{Chirikov, B. V., Izrailev, F. M., Shepelyanskii, D. L.,  {\it Quantum chaos: Localization vs. ergodicity}. Phys.  D 33 (1988), n. 1-3, 77-88.}

{\color{black}
\bibitem{deGo04}
de Gosson, M. A. {\it On the notion of phase in mechanics}. J. Phys. A: Math. Gen. 37 (2004), 7297-7314
}

\bibitem{Wiener}
Della Riccia, G.; Wiener, N. {\it Wave mechanics in classical phase space, Brownian motion, and quantum theory}. {J. Math. Phys.} 6 (1966), 1372-1383

\bibitem{Dodin1}
Dodin, I.Y. {\it Geometric view on noneikonal waves}. Phys. Lett. A 378 (2014) 1598-1621

\bibitem{Dodin20}
 Dodin, I.Y.; Startsev, E.A. {\it  On applications of quantum computing to plasma simulations}. {\tt arXiv:2005.14369}
 
\bibitem{Engel19}
Engel, A.; Smith, G.; Parker, S.E. {\it  Quantum algorithm for the Vlasov equation}. Phys. Rev. A 100 (2019), n. 6, 062315 

\bibitem{Engel20}
Engel, A.; Smith, G.; Parker, S.E. {\it A framework for applying quantum computation to nonlinear dynamical systems} {\tt arXiv:2012.06681}


\bibitem{Faure}
Faure, F. {\it Prequantum chaos: Resonances of the prequantum cat map}. J. Mod. Dyn. 1 (2007), n. 2, 255-285


\bibitem{FoHoTr19}
Foskett, M.S.; Holm, D.D.; Tronci, C. 
\newblock {\it Geometry of nonadiabatic quantum hydrodynamics}. { Acta Appl. Math.} 162 (2019), 1-41

\bibitem{Frenkel}
Frenkel, J. Wave Mechanics; Advanced General Theory. Claredon Press. Oxford. 1934.

\bibitem{Gaitan20}
{Gaitan, F. {\it Finding flows of a Navier-Stokes fluid through quantum computing}, npj Quantum Inf. 6 (2020), 61}

\bibitem{GBTr20}
Gay-Balmaz, F.; Tronci, C. {\it Madelung transform and probability densities in hybrid quantum-classical dynamics}. Nonlinearity, 33 (2019), n. 10, 5383-5424

\bibitem{GBTr21a}
Gay-Balmaz, F.; Tronci, C. {\it  From quantum hydrodynamics to Koopman wavefunctions I}. Lecture Notes in Comput. Sci. (to appear). {\tt arXiv:2104.13185}


\bibitem{Giannakis}
Giannakis, D.; Ourmazd, A.; Slawinska, J.; Schumacher, J. {\it Quantum compiler for classical dynamical systems}. {\tt arXiv:2012.06097}


\bibitem{GBTr21b}
Tronci, C.; Gay-Balmaz, F. {\it  From quantum hydrodynamics to Koopman wavefunctions II}.  Lecture Notes in Comput. Sci. (to appear). {\tt arXiv:2104.13172}

\bibitem{GuSt80}
Guillemin, V., Sternberg, S. {\it The moment map and collective motion}. Ann. Phys. 127 (1980), 220-253 

\bibitem{HoKu}
Holm, D.D.; Kupershmidt, B.A. {\it Poisson brackets and Clebsch representations for magnetohydrodynamics, multifluid plasmas, and elasticity}. Phys. D 6 (1983), 347-363

\bibitem{HoMaRa1998}
Holm, D.D.; Marsden, J.E.;  Ratiu, T.S. {\it The Euler--Poincar{\'e} equations and semidirect products with applications to continuum theories}. {Adv. Math.}, 137 (1998), 1-81

\bibitem{Sugny}
Jauslin, H.R.; Sugny, D. {\it Dynamics of mixed quantum--classical systems, geometric quantization and coherent states}. In ``Mathematical Horizons for Quantum Physics''.  Lect. Notes Ser. Inst. Math. Sci. Natl. Univ. Singap., 20, 65-96. World Scientific. 2010.


\bibitem{Joseph20}
Joseph, I. {\it Koopman-von Neumann approach to quantum simulation of nonlinear classical dynamics}. Phys. Rev. Res. 2 (2020), n. 4, 043102

\bibitem{Kirillov}
Kirillov, A.A. {\it Geometric quantization}. In ``Dynamical Systems IV'', 139-176, Encyclopaedia Math. Sci., 4, Springer, 2001.


\bibitem{Koopman}
Koopman, B.O. {\it Hamiltonian systems and transformations in Hilbert space}. Proc. Nat. Acad. Sci. 17 (1931), 315
 
\bibitem{Ko1970}
Kostant, B. {\it Quantization and unitary representations}, In ``Lectures in modern analysis and applications III'', 87--208. Lecture Notes in Math. 170, Springer, 1970

\bibitem{Liu20}
Liu, J.-P.; Kolden, H. \O.; Krovi, H. K.; Loureiro, N. F.; Trivisa, K.; Childs, A. M.; {\it Efficient quantum algorithm for dissipative nonlinear differential equations},	{\tt arXiv:2011.03185}

\bibitem{MaRa}
Marsden, J.E.; Ratiu, T.S. Introduction to Mechanics and Symmetry. Springer. 1998

\bibitem{MaWe2}
Marsden, J.E.; Weinstein, A. {\it  Coadjoint orbits, vortices, and Clebsch variables for incompressible fluids.}  Phys. D  7 (1983), 305-323

\bibitem{MaWe1}
Marsden, J.E.; Weinstein, A. {\it The Hamiltonian structure of the
Maxwell-Vlasov equations.}  Phys. D  4  (1981/82), no. 3, 394--406

\bibitem{MaWeRaScSp}
Marsden, J.E.; Weinstein, A.; Ratiu, T.; Schimd, R.; Spencer, R.G.
{\it  Hamiltonian systems with symmetry, coadjoint orbits and plasma
physics},  Atti Accad. Sci. Torino Cl. Sci. Fis. Mat. Natur.  117
(1983), no. 1, 289--340

\bibitem{Mauro}
Mauro, D. {\it On Koopman-von Neumann waves}. {Int. J. Mod. Phys. A} 17 (2002), 1301

\bibitem{Morrison1}
Morrison, P.J. {\it Hamiltonian field description of two-dimensional
vortex fluids and guiding center plasmas.} Princeton
Plasma Physics Laboratory Report, PPPL-1783 (1981).

\bibitem{Morrison2bis}
Morrison, P.J. {\it The Maxwell-Vlasov equations as a continuous
Hamiltonian system.} Phys. Lett. A 80 (1986), no. 5--6, 383--386

\bibitem{NielsenChuangBook}
Nielsen,  M.A.; Chuang, I.L. {Quantum Computation and Quantum Information.} Cambridge University Press 2010.

\bibitem{Neiss}
Neiss, R.A. {\it Generalized symplectization of Vlasov dynamics and application to the Vlasov-Poisson system}. Arch. Rational Mech. Anal. 231 (2019), n. 1, 115-151



\bibitem{DOE}
Shenkel, T.; Dorland, B; Baczewski, A.; Boshier, M.; Collins, G.; Dubois, J.; Houck, A.; Humble, T.; Loureiro, N.; Monroe, C. {\it Fusion Energy Sciences Roundtable on Quantum Information Science}. Department of Energy Report. May 01-02, 2018. United States.

\bibitem{Squire}
Squire, J.; Qin, H.; Tang, W. M.; Chandre, C. {\it The Hamiltonian structure and Euler-Poincar\'e formulation of the Vlasov-Maxwell and gyrokinetic systems}. Phys. Plasmas 20 (2013), n. 2, 022501 

\bibitem{Sudarshan}
Sudarshan, E.C.G. {\it Interaction between classical and quantum systems and the measurement of quantum observables}. {Pr\={a}ma\d{n}a} 6 (1976), 117

\bibitem{tHooft}
't Hooft, G. {\it Quantummechanical behaviour in a deterministic model}. {Found. Phys. Lett.} 10 (1997), 105-111

\bibitem{TronciRec}
Tronci, C.{\it A Lagrangian kinetic model for collisionless magnetic reconnection}. Plasma Phys. Control. Fusion. 55 (2013), n. 3, 035001 


\bibitem{VanHove}
van Hove, L. On certain unitary representations of an infinite group of transformations. PhD Thesis (1951). Word Scientific 2001

\bibitem{VonNeumann2}
von Neumann, J. {\it Zur Operatorenmethode in der klassischen Mechanik}. {Ann. Math.} 33 (1932). 587-642
 

\end{thebibliography}
\end{document}